# Short-range spin order and frustrated magnetism in $Mn_2InSbO_6$ and $Mn_2ScSbO_6$.

Sergey Ivanov[a,c], Per Nordblad[b], Roland Mathieu[b], Roland Tellgren[c], Ekaterina Politova[a], Gilles André[d].

a-Department of Inorganic Materials, Karpov' Institute of Physical Chemistry, Vorontsovo pole,10 105064, Moscow K-64, Russia
b-Department of Engineering Sciences, Uppsala University, Box 534, SE-751 21, Uppsala, Sweden
c-Department of Materials Chemistry, Uppsala University, Box 534, SE-751 21 Uppsala, Sweden
d-Laboratoire Leon Brillouin, CEA, Saclay, France

**Abstract**

The complex metal oxides $Mn_2ASbO_6$ (A=In, Sc) with a corundum-related structure $A_3BO_6$ have been prepared as polycrystalline powders by a solid state reaction route. The crystal structure and magnetic properties have been investigated using a combination of X-ray and neutron powder diffraction, electron microscopy, calorimetric and magnetic measurements. At room temperature these compounds adopt a trigonal structure, space group $R\bar{3}$ with $a$ = 8.9313(1) Å, $c$ = 10.7071(2) Å (for In) and $a$ = 8.8836(1) Å, $c$ = 10.6168(2) Å (for Sc) which persists down to 1.6 K. The Mn and A cations were found to be randomly distributed over the A-sites. The overall behavior of the magnetization of $Mn_2InSbO_6$ and $Mn_2ScSbO_6$ is quite similar. In spite of the relatively large amount of Mn ions on the A-site, only short-ranged magnetism is observed. Neutron diffraction patterns of $Mn_2InSbO_6$ showed no evidence of a long-range magnetic ordering at 1.6 K, instead only a weak diffuse magnetic peak was observed at low temperatures. The factors governing the observed structural and magnetic properties of $Mn_2ASbO_6$ are discussed and compared with those of other Mn-containing complex metal oxides with a corundum-related structure. The influence of the A-cation sublattice on magnetic properties is also considered.

**Keywords:** magnetic oxides, multiferroics, neutron scattering, crystal structure, magnetic properties, ferroelectricity

# Introduction

The origin and understanding of coupling phenomena between different physical properties within one material is a central subject of solid state science. A great deal of theoretical and experimental attention in this field is currently focused on the coupling between magnetism and ferroelectricity, as can be encountered in the so-called multiferroics [1, 2]. These compounds present opportunities for a wide range of potential applications [3, 4] in addition to that the fundamental physics of multiferroic materials is rich and fascinating [5-8]. The coexistence of ferromagnetism and ferroelectricity is difficult to achieve for many reasons [2, 9] and only very few multiferroic materials are known [5]. First, and most fundamentally, the cations which are responsible for the electric polarization in conventional ferroelectrics have a formally empty *d*-electron configuration. In contrast, ferromagnetism requires unpaired electrons, which in many materials are provided by *d* electrons of transition metal ions. Therefore the coexistence of the two phenomena, although not prohibited by any physical law or symmetry consideration, is discouraged by the local chemistry that favors one or the other but not both. In practice, alternative mechanisms for introducing the polar ion displacements and spin ordering are still needed. Theory gives us a good guide in what type of materials we can expect a large coupling effect. There are several important conclusions that we can draw regarding future directions of multiferroic research [2, 5, 9]. However, experimental activity has been limited due to a lack of novel materials. Existing multiferroic compounds belong to different crystallographic classes, and although some general rules governing their behaviour are already established [10, 11], there is as yet no complete or general understanding of the origin of multiferroic behaviour.

With a few exceptions, the multiferroics materials that have been investigated are transition metal oxides, mainly with a perovskite-related structure. The unique range of responses, coupled with the flexibility of perovskites in accommodating a broad spectrum of atomic substitutions, provides a robust platform for probing correlations between structure, bulk chemistry and magnetoelectric properties. But another key question is in what kinds of other structural families we can also expect multiferroic behaviour including different types of magnetic and electric interactions which are responsible for the magnetic and ferroelectric properties and the magnetoelectric coupling. The construction of new inorganic compounds through the rational combination of different cations has become an area of great interest due to their potential modern applications.

In order to obtain new frameworks, we choose $Mn^{2+}$ compounds with a corundum-related structure due to the octahedral coordination preferences of this cation. The capability of corundum-structured oxides to substitute one ion with another (or a set of ions) without remarkable structural distortion provides a way of obtaining new magnetoelectric materials. In an attempt to open up a new avenue of multiferroic materials, we have investigated the compounds with corundum-related structure $A_3BO_6$ ($A$=Mn, In, Sc; $B$=Te,Sb). Several $A_3TeO_6$ ($A$=Mn, Co, Ni, Cu) have only recently been reported as new potential multiferroics [12-14]. $Mn_3TeO_6$ crystallizes pseudocubic; in a rhombohedral structure with $\alpha_R$ near 90º and has fairly close packing of oxygen atoms [15]. Since the oxide chemistry of tellurium shows a striking relationship to that of antimony [16, 17], attempts were made to synthesize complex metal oxides of this structural type in which $Te^{6+}$ is completely substituted by $Sb^{5+}$.

In this article we focused on $Mn_2ASbO_6$ (A =In, Sc) because these compounds have been known as ferroelectrics [18, 19]. The literature data on $Mn_2InSbO_6$ (MISO) and $Mn_2ScSbO_6$ (MSSO) are very scarce. The MISO compound was first prepared as a powder by *Bayer* in [20, 21] but detailed investigations of this system has been mainly focused on crystallographic aspects. Later, the solid state reaction performed by *Kosse* in [18, 19] gave a single-phase ceramic MISO and this compound was described as antiferroelectric with $T_c$=485 K. MSSO was also prepared by *Kosse* in [18, 19] and similar dielectric anomaly was registered around 470 K. Existing experimental data are very limited and detailed understanding of these dielectric anomalies is still an open question. But it seems clear that the dielectric properties of the obtained phases can be affected by the polarizability of the transition metal and by the extent of the structural distortion of the $AO_6$ octahedra. It is also necessary to note that the structural and magnetic properties for both compounds not yet have been clarified.

We have studied the structural and magnetic properties of $Mn_2ASbO_6$ to determine possible differences in their magnetic properties, arising because of differences between $Mn^{2+}$ and $A^{3+}$ ions. In this paper we report the results of X-ray and neutron diffraction studies and magnetic measurements of these compounds which clearly show the absence of long-range magnetic order. As an important supplement, the regularities of structure and magnetic properties as a function of the $A^{3+}$ cation were considered. In the complex manganese oxides, the introduction of other elements, which exhibit dissimilar electronic configuration to Mn, should lead to dramatic effects associated with the electronic configuration mismatch between Mn and the substituted ions. In this sense, and in order to avoid the remarkable lattice distortions, the In and Sc cations have been selected. When several crystallographic sites contain mixtures of two (or more) different cations, control of the positional ordering of the cations can provide an additional tool for mediating the spin and dipole ordering.

**Results**

According to the elemental analyses done on 20 different crystallites, the metal composition of MISO and MSSO are $Mn_{1.97(4)}In_{1.02(3)}Sb_{1.01(3)}$ and $Mn_{1.99(4)}Sc_{0.98(3)}Sb_{1.03(3)}$ if the sum of the cations is assumed to be 4. The oxygen content determined with an iodometric titration for the samples was determined as 5.98(3) and 5.99(3). All these values are very close to the expected ratios and permit to conclude that the sample compositions are the nominal ones. The microstructure of the obtained powders, observed by scanning electron microscopy, reveals uniform and fine grain distribution.

Second harmonic generation (SHG) measurements at room temperature for both compositions gave a negative result, thus testifying that at this temperature the MISO and MSSO compounds probably possess a centrosymmetric crystal structure. These samples still could be non-centrosymmetric, but at a level detectable only with sensitivities beyond $10^{-2}$ of quartz [22].

The first crystallographic characterization of the samples was performed by X-ray powder diffraction analysis at room temperature which showed that the prepared samples of MISO and MSSO formed single phase powders. The room temperature XRD patterns could be indexed on the basis of a hexagonal unit cell with $a$ = 8.9305(1) Å and $c$=10.7066(2) Å (for MISO) and $a$ = 8.8836(1) Å and $c$ =10.6168(2) Å (for MSSO).

These values are in a reasonable agreement with the lattice parameters obtained by *Kosse* in [19] and *Bayer* in [21]. The XRPD patterns could be successfully refined by the Rietveld method using s. g. $R\bar{3}$. It was found that $Mn^{2+}$ and $In^{3+}$, having a similar ionic sizes ($Mn^{2+}$=0.83 Å and $In^{3+}$=0.80 Å) prefer to occupy the same 18f position. A possible anti-site disorder in the A-sublattice of the sample has also been investigated but on the basis of the X-ray data it is difficult to distinguish between $In^{3+}$ and $Sb^{5+}$ cations due to their similar scattering power. The obtained results for MSSO show that there is no measurable anti-site disorder between the In and Sc sublattices. The Mn/A-O and Sb-O bond lengths calculated from the refined lattice parameters and atomic coordinates are in good agreement with earlier observed for another Mn-based antimonates [23, 24]. Furthermore, the corresponding bond valence sum calculations are consistent with the presence of $Mn^{2+}$, $In^{3+}$, $Sc^{3+}$, $Sb^{5+}$ and $O^{2-}$ ions. The final results from the Rietveld refinements with observed, calculated and difference plots for XRPD patterns of MISO and MSSO at 295 K are shown in figures 1a and 1b and the structure in figure 2.

The upper panels of figure 3 shows the low-temperature magnetic susceptibility of MISO and MSSO recorded in large (3 Tesla, main frames) and low (5 mT, insets) magnetic fields. Looking first at the main frames, it appears that the overall behavior of the magnetization of both systems is quite similar, with the magnetization monotonously decreasing as the temperature increases. Yet an inflection point can be seen in the magnetization curves below 25K, as well as a tiny irreversibility. This suggests that some short-range ordered magnetism contributes to the magnetization at low temperatures. As seen in the insets of the upper panels the low-field magnetization is instead dominated by weak spontaneous magnetic moments contributing to the measured magnetization below 60K, possibly associated with an impurity phase of an amount far below the detection limit of the XRD and NPD experiments. The low-field zero field-cooled magnetization curves of both systems display cusps below 10K, which looks spin-glass like.

The behavior of the magnetic susceptibility of MISO and MSSO at higher temperatures is shown in the lower panels of figure 3. As seen in the insets of the lower panels of Figure 2, the susceptibility data of MISO and MSSO above 25K follows a Curie-Weiss law M/H(T)=C/(T+θ) with values of θ ~ -64 K and -76K, and effective Bohr magneton number p of 5.87 $\mu_B$ and 5.79 $\mu_B$ respectively. The p values are in very good agreement with the theoretically expected value for $Mn^{2+}$ in a $3d^5$ configuration (S=5/2, L=0, p=5.92 $\mu_B$). The relatively large |θ| value indicates a significant magnetic interaction and frustration in both compounds.

The temperature dependence of the heat capacity, C, of MISO and MSSO is plotted in figure 4. No distinct features can be observed around 50-60K, i.e. near the temperature below which the low-field magnetization sharply increases. On the other hand, a broad maximum is observed in the C(T)/T and a corresponding bump in the C(T) data (see main frame and inset of figure 4. This broad maximum reflects the short-range magnetic order suggested by the magnetization measurements. It thus seems that the magnetic frustration present in MISO and MSSO hinders any long-ranged antiferromagnetic order.

We started to refine the crystal structure of the MISO sample using NPD data at 60 K (above the inflection point that can be seen in the magnetization curves). To test the space group found from the refinements of the XRD data several centrosymmetric trigonal space groups were initially considered. Rietveld refinements were carried out in all space groups, but a clearly superior fit was obtained using space group $R\bar{3}$ including the structural model refined previously from single crystal XRD data. The obtained cation positions and occupation factors for Mn and In are very similar, but we were able to determine more accurately the oxygen positions due to the characteristics of the neutron scattering. No vacancies were observed in the cationic or in the anionic substructures. Accordingly, the chemical composition seems to be very close to the nominal one and therefore, the Mn oxidation state can be assumed to be +2.

Table 1 lists the structural parameters for MISO and MSSO: atomic coordinates, lattice parameters, thermal displacement parameters, as well as the reliability factors. Selected bond lengths are listed in Table 2. The NPD patterns of MISO at 1.6 K and 60K are depicted in Figure 5. Polyhedral analysis of the $Mn^{2+}$, $In^{3+}$, $Sc^{3+}$ and $Sb^{5+}$ cations for MISO and MSSO at room temperature is presented in Table 3. The crystal structure of investigated samples is similar to the prototype $Mg_3TeO_6$ and can be derived from a close packing of strongly distorted hexagonal oxygen layers parallel to (001), with Mn/A and two distinct Sb atoms in the octahedral interstices (see figure 2). Both $SbO_6$ octahedra exhibit 3-fold symmetry and are fairly regular (see Table 2), with an average Sb-O distance, which is in good agreement with the average Sb-O distances of other Mn antimonates [23, 24]. Each $SbO_6$ octahedron shares edges with six $(Mn/A)O_6$ octahedra but none with other $SbO_6$ octahedra. The $(Mn/A)O_6$ octahedron is considerably distorted which is reflected by the variation of the Mn/A-O distances (Table 2). Each $(Mn/A)O_6$ octahedron shares four edges with adjacent $(Mn/A)O_6$ octahedra, one edge with a $Sb(1)O_6$ and another edge with the $Sb(2)O_6$ octahedron. The shared edges of the $SbO_6$ octahedra have somewhat shorter O-O distances than the non-shared edges. This is may be due to the high valence of antimony, which has the tendency to keep as far as possible from the mixed $Mn^{2+}/A^{3+}$ cations. Each of the two crystallographically independent O atoms is coordinated by one Sb and three Mn/A atoms in a distorted tetrahedral manner. In addition to the octahedral holes (x,y,z), which are occupied by Mn/A there exists an identical number of similar structural voids (x,y,z+0.5), which also could be filled by substituted A-cations. This hypothetical arrangement gave a significantly worse fit of our diffraction data and this model was rejected.

A careful look at NPD patterns collected at 1.6K and 60K for MISO (see figure 5) shows that a weak broad magnetic peak (diffuse scattering) is present at 1.6 K in addition to the nuclear Bragg peaks. This suggests the existence of short-range magnetic correlations, however with a significant fraction of the spins remaining disordered. Additional pure magnetic maxima in the 1.6 K neutron diffraction data, expected for a three-dimensionally ordered antiferromagnetic state were not observed, confirming the absence of long-range magnetic order. Position and profile parameters of this broad diffuse halo were obtained by peak profile fitting. The width of the broad diffuse peak has been used for an estimation of the correlation length at 1.6 K. If we picture the short-range ordered magnetic structure in MISO as a simple alternation of ordered and disordered domains, the domain size can be estimated using the *Scherrer* formula [25] based on the reciprocal relation between cluster size $D(Å)$ and peak half-width broadening *FWHM* (degrees) according to

$$D = \lambda \; x \; 57.3 / \cos\Theta \; x \; FWHM$$

By taking the actual experimental parameters (after correction for instrumental resolution) the calculation results in a characteristic value of 10 Å for magnetic cluster. The counting time for current measurement was insufficient to give data of adequate quality to provide a detailed analysis of the short-range order here. However, it is useful to note that the nearest neighbour Mn-Mn distance in this structure is 3.24 Å, while the next-nearest neighbour distance is 4.45 Å, showing the broad peak does not correspond directly to a Mn-Mn correlation. This is near to (101) and (110) reflection positions, suggesting that the broad peak could be related to magnetic correlations associated with these Bragg reflections. The lack of long-range magnetic order in MISO may be related to a random distribution of Mn and In cations on the A-sites for corundum-related structure. This mixture of cations promotes magnetic frustration, which hinders long range magnetic order to establish.

**Discussion**

Crystal chemistry of tellurium shows that $Te^{6+}$ is very similar in its behavior to the isoelectronic $Sb^{5+}$ (both cations are $nd^{10}$-ions). Both have a comparable ionic radius ($Te^{6+}$=0.56 Å, $Sb^{5+}$=0.60 Å), the same strong preference for octahedral oxygen coordination and both can be substituted in oxide compounds, e.g. of the rutile, ilmenite and perovskite types of structure [26-28]. If one takes into account the values of the polyhedra distortions (see Table 3), possible ferroic properties of $Mn_2ASbO_6$ may be connected with the strongly distorted A- sublattice, where the cations are significant displaced from the polyhedral centers. It is worth to notice that the volumes of the $Sb_1$ and $Sb_2$ polyhedra are smaller in comparison with the octahedral $(Mn/A)O_6$.

The Mn and In (or Sc) cations are apparently disordered over the octahedral *A*-site of the corundum-related structure with *2/3* of the site occupied by a magnetic cation. With such a high concentration of magnetic species, well above the percolation limit, it is natural to expect that the compound order antiferromagnetically at low temperatures. We would expect the superexchange along a pathway Mn-O-Mn to be the strongest magnetic interaction in such a system, and that a G-type magnetic structure would be adopted. However, the behavior of the FC and ZFC susceptibilities and the absence of magnetic scattering from the low-temperature diffraction pattern all indicate that this is not the case. The absence of magnetic Bragg peaks proves that there is no long-range magnetic ordering. Both MISO and MSSO exhibit frustrated magnetism. From the Curie-Weiss analysis and resulting θ temperatures, the magnetic frustration may be slightly larger in MSSO. The magnetic frustration is also large in $Mn_3TeO_6$ [29], albeit in that case a long-ranged antiferromagnetic order is established below 24K in spite of the frustration. We therefore conclude that we are dealing with a short-range magnetic orderings and this conclusion is supported by the magnetic susceptibility and heat capacity data.

The principal geometric parameters for qualitative understanding of magnetic interactions are the Mn-Mn distances and the Mn-O-Mn angles. For superexchange interactions the most favorable metal-oxygen-metal angle is 180°. However, in the same time the near 90° angle is most unfavorable for a superexchange interaction. According to the *Goodenough - Kanamori* rules [30,31], a decreasing Mn-O-Mn angle results in a monotonic decrease from a large antiferromagnetic coupling at an 180° bond to a weak ferromagnetic coupling at a 90° bond. In the studied structural family Mn cations order in alternating layers along the trigonal axis which are rather isolated from one another, and no 180° connectivity (as in the case of perovskite structure) is to be expected. Indeed, in the case of corundum-related structure the bonding angles of Mn-O-Mn are between 94° and 117°, which causes a perfect geometrical frustration and negligible super-exchange interactions [32].

The substitution of Mn by In, Sc ions appreciably affects the unit cell parameters (see Table 4). The relative variation of $a$ and $c$ is clearly related to the ionic radius of the substituting ions: both $a$ and $c$ show a linear decrease with doping by $In^{3+}$ and $Sc^{3+}$, which have a radius smaller than that of $Mn^{2+}$.

Doping of $Mn_3TeO_6$ with ions of different sizes and valence states would affect the polarization and electrical properties. Remembering that the polarizability may be estimated as the ratio "valence/ionic radius" we can calculate this value for our cations (Mn=2.41, Sc=4, In=3.75, Sb =8.33, Te=10.7). $Mn^{2+}$ has a larger ionic radius than $In^{3+}$ and $Sc^{3+}$, therefore, when $In^{3+}$ and $Sc^{3+}$ ions are substituted for $Mn^{2+}$ ions, the cage of the structure would have more ''rattling space'' and enhanced polarization. The space for the A-cation would then increase, and the A-cations can be polarized more easily.
In the same time, during Sb substitution the polarizability of the B-cations are significantly smaller. Thus, antiferroelectric properties as a competition of two factors are expected to be slightly worse with doping.
Substitution of the transition element leads to decrease in the dielectric permittivity ε max and in the dielectric loss tan δ max values, measured at frequency f=1 kHz, from ε max =8000 and tan δ max =10 for $Mn_3TeO_6$ to ε max =900 and tan δ max =7 for MSSO and ε max =2400 and tan δ max =6 for MISO.
The observed changes in the dielectric properties evidently indicate influence of materials conductivity governed by the amount of the transition element (manganese), though a contribution of the microstructure parameters may take place as well [33]. Other factors as the conductivity or domain size and distribution can also affect the final properties of a ferroelectric ceramic.

The observed degradation of the magnetic properties with a substitution of Mn-Te on In-Sb (or Sc-Sb) suggests that these combinations of substituted cations significantly modify the exchange interactions. It is clear since the In and Sc ions have preference to occupy *A*-sites only and the Mn-sublattice gets diluted with the increase of their concentration. Finally, the magnetic exchange interactions in the Mn-sublattice get weakened with every step of substitution. This explains the pronounced degradation of the magnetic properties.

**Conclusions**

$Mn_2AsbO_6$ (A=In, Sc) have been identified and found to crystallize in the corundum-type structure. Mixed Mn/A sites provide a random distribution of Mn–Mn distances. In spite of the relatively large amount of magnetic Mn ions on the A-site in $Mn_2AsbO_6$, no evidence of long-range magnetic order was obtained from neutron powder diffraction (or from macroscopic magnetization and heat capacity measurements). The appearance of diffuse scattering at low temperatures indicates that a significant fraction of the spins remains disordered at 1.6K. Our results suggest that the random distribution of Mn–Mn magnetic interactions with partly (2/3) occupied Mn sites and the associated magnetic frustration are responsible for the observed lack of long-range magnetic order.

**Experimental Section**

High quality ceramic samples of MISO and MSSO were prepared by a conventional solid state reaction of inorganic precursor compounds in air. Starting materials were $MnCO_3$, $In_2O_3$, $Sc_2O_3$ and $Sb_2O_5$. The reactive powders were mechanically ground and mixed in an agate mortar in an appropriate stoichiometric ratio. The mixtures were calcined in a platinum crucible at 1070 K for 4 h. After grinding at room temperature they were annealed for 14 h at 1223 K. The annealing was repeated at 1323 K and 1423 K during 14 h. For synthesis in air the mixtures were quenched to room temperature. The mixtures were heated to 1573 K with a slope of 10 K/min, annealed for 14 h. and slowly cooled to room temperature. In air MISO and MSSO could be prepared as pure, single-phase ceramics with a corundum-related structure without undesirable stable impurity phases. The purity of the powder samples was checked from powder X-ray diffraction (XRPD) patterns obtained with a high-resolution SIEMENS D-5000 diffractometer. Filtered Cu K$\alpha$ radiation ( $\lambda = 1.54056$ Å ) was used. The diffraction diagram was measured from 10° to 140° in $2\theta$ with step size 0.02° ($2\theta$) and 15 s counting time per step. Additionally, the low angle part of the pattern for MISO was measured with longer, 20 s counting time and shorter, 0.01° step size that allowed for additional analysis of the background. No clear deviation from the rhombohedral symmetry and extra impuriry reflections could be observed by XRPD.

The chemical compositions of the prepared crystals and ceramic samples were determined by energy-dispersive spectroscopy (EDS) using a JEOL 840A scanning electron microscope and INCA 4.07 (Oxford Instruments) software. The analyses performed on several samples showed that the concentration ratios of Mn: A: Sb were as expected within the instrumental resolution (0.05).

Polycrystalline materials were characterized by SHG measurements in reflection geometry, using a pulsed Nd:YAG laser ($\lambda$=1.064 µm). The SHG signal $I_{2\omega}$ from the sample was measured relative to an $\alpha$-quartz standard at room temperature in the Q-switching mode with a repetition rate of 4 Hz.

The magnetization experiments were performed in a Quantum Design MPMSXL 5 T SQUID magnetometer. The magnetization (M) was recorded as a function of temperature T in the interval 5-300 K in 5 mT (50 Oe) and 3 T field using zero-field-cooled (ZFC) and field-cooled (FC) protocols.

Specific heat measurements were performed using a relaxation method between 2 K and 70 K on a Physical Properties Measurement System (PPMS6000) from Quantum Design Inc.

Because the neutron scattering lengths of Mn and In are very different, the chemical composition and cation distribution between different crystallographic sites can be observed by neutron powder diffraction (NPD) with good precision ($b_{Mn} = -3.75$, $b_{In} = 4.07$, $b_{Sb} = 5.57$ fm). The neutron scattering length of oxygen ($b_O = 5.80$ fm) is comparable to those of the heavy atoms and NPD provide an accurate information on its position and stoichiometry. Neutron diffraction study was carried out using the G4.1 diffractometer ($\lambda = 2.4255$ Å) at LLB, Saclay, France. NPD patterns were registered at temperatures below (1.6K) and above (60K) a possible magnetic phase transition. Structural refinements were performed by the Rietveld method, using the *FULLPROF* software [34]. The diffraction peaks were described by a pseudo-Voigt profile function, with a Lorentzian contribution to the Gaussian peak shape. A peak asymmetry correction was made for angles below 35° in 2θ. Background intensities were estimated by interpolating between 40 selected points (NPD experimental data at 1.6 K) or described by a polynomial with six coefficients (NPD experimental data at 60 K). The variant for which the structural refinement was stable and the reliability factors at a minimum was chosen as the final model. Analysis of the coordination polyhedra of the cations was performed using the *IVTON* software [35]. The structure of $Mn_3TeO_6$ (space group R-3) [15] was taken as a starting model for the Rietveld refinements. Special software for profile fitting (program *PROFIT* [36]) was used for analysis of the broad diffuse peak. The program fits pre-selectable profile functions, *e.g.* Gaussian or Lorentzian, into measured peaks or peak clusters and individually refines sets of three profile parameters each for peak position 2Θ, half-width *FWHM*, and peak height *H*. This procedure permits the separation of broad diffuse magnetic scattering contributions from the pure Bragg peaks and thus the separation of disorder and long-range order effects, respectively.

**Acknowledgements**

Financial support of this research from the Royal Swedish Academy of Sciences, the Swedish Research Council (VR), the Göran Gustafsson Foundation and the Russian Foundation for Basic Research is gratefully acknowledged. We also gratefully acknowledge the support from N. Sadovskaya and S. Stefanovich during the EDS cation analysis and the second harmonic generation testing.

Table 1 Summary of the results of the structural refinements of the $Mn_2InSbO_6$ and $Mn_2ScSbO_6$ samples using XRPD and NPD data.

| Composition | $Mn_2InSbO_6$ | | | $Mn_2ScSbO_6$ |
|---|---|---|---|---|
| Experiment | XRPD | NPD | NPD | XRPD |
| T,K | 295 | 60 | 1.6 | 295 |
| a[Å] | 8.9313(2) | 8.9292(4) | 8.9296(4) | 8.8836(1) |
| c[Å] | 10.7071(3) | 10.7031(5) | 10.7041(5) | 10.6168(2) |
| s.g. | $R\bar{3}$ | $R\bar{3}$ | $R\bar{3}$ | $R\bar{3}$ |
| | **Mn/In** | | | **Mn/Sc** |
| $x$ | 0.0398(3) | 0.0381(6) | 0.0372(6) | 0.0391(2) |
| $y$ | 0.2651(2) | 0.2536(7) | 0.2599(7) | 0.2645(2) |
| $z$ | 0.2112(3) | 0.2059(8) | 0.2045(8) | 0.2121(3) |
| $B[Å]^2$ | 0.47(1) | 0.38(4) | 0.35(4) | 0.39(1) |
| | **Sb1** | | | |
| $x$ | 0 | 0 | 0 | 0 |
| $y$ | 0 | 0 | 0 | 0 |
| $z$ | 0 | 0 | 0 | 0 |
| $B[Å]^2$ | 0.28(1) | 0.22(4) | 0.19(4) | 0.31(1) |
| | **Sb2** | | | |
| $x$ | 0 | 0 | 0 | 0 |
| $y$ | 0 | 0 | 0 | 0 |
| $z$ | 0.5 | 0.5 | 0.5 | 0.5 |
| $B[Å]^2$ | 0.34(1) | 0.27(4) | 0.24(4) | 0.36(1) |
| **O1** | | | | |
| $x$ | 0.0251(5) | 0.0258(6) | 0.0226(4) | 0.0281(7) |
| $y$ | 0.2008(6) | 0.1961(7) | 0.1955(6) | 0.2028(6) |
| $z$ | 0.3970(4) | 0.3976(5) | 0.3973(5) | 0.3949(5) |
| $B[Å]^2$ | 0.57(2) | 0.46(5) | 0.42(5) | 0.61(2) |
| **O2** | | | | |
| $x$ | 0.1832(4) | 0.1881(5) | 0.1871(5) | 0.1849(5) |
| $y$ | 0.1568(6) | 0.1611(4) | 0.1597(4) | 0.1549(7) |
| $z$ | 0.1163(5) | 0.1114(6) | 0.1108(6) | 0.1216(5) |
| $B[Å]^2$ | 0.66(2) | 0.50(5) | 0.46(5) | 0.71(2) |
| $R_p$,% | 4.58 | 4.14 | 5.16 | 4.79 |
| $R_{wp}$,% | 5.73 | 5.51 | 6.42 | 5.94 |
| $R_B$(%) | 4.14 | 2.84 | 3.19 | 4.32 |
| $\chi^2$ | 1.97 | 2.17 | 1.79 | 2.09 |

Table 2. Selected bond lengths [Å] of the Mn$_2$InSbO$_6$ and Mn$_2$ScSbO$_6$ samples.

| Bonds, Å | | | Mn$_2$InSbO$_6$ | | Mn$_2$ScSbO$_6$ |
|---|---|---|---|---|---|
| T,K | | | 1.6 | 295 | 295 |
| Mn/B | | O1 | 2.048(6) | 2.056(6) | 2.006(5) |
| | | O1 | 2.352(4) | 2.363(6) | 2.379(6) |
| | | O1 | 2.240(4) | 2.248(6) | 2.222(6) |
| | | O2 | 2.194(6) | 2.202(7) | 2.155(7) |
| | | O2 | 2.209(4) | 2.218(5) | 2.193(7) |
| | | O2 | 2.133(7) | 2.145(7) | 2.126(7) |
| Sb1 | | O2 | 1.974(3) | 1.989(4) | 1.999(5) |
| Sb2 | | O1 | 2.020(4) | 2.033(5) | 2.025(4) |

Table 3. Polyhedral analysis of Mn$_2$InSbO$_6$ and Mn$_2$ScSbO$_6$ at room temperature (cn - coordination number, x – shift from centroid, ξ- average bond distance with a standard deviation, V- polyhedral volume, ω- polyhedral volume distortion.

Mn$_2$InSbO$_6$ T=295K space group $R\overline{3}$.

| Cation | cn | x(Å) | ξ (Å) | V(Å$^3$) | ω | Valence |
|---|---|---|---|---|---|---|
| Mn/In | 6 | 0.054 | 2.205+/-0.103 | 12.7(1) | 0.103 | 2.02/2.73 |
| Sb1 | 6 | 0 | 1.974+/-0.004 | 10.1(1) | 0.013 | 5.19 |
| Sb2 | 6 | 0 | 2.020+/-0.005 | 10.9(1) | 0.004 | 5.28 |

Mn$_2$ScSbO$_6$ T=295K space group $R\overline{3}$.

| Cation | cn | x(Å) | ξ (Å) | V(Å$^3$) | ω | Valence |
|---|---|---|---|---|---|---|
| Mn/Sc | 6 | 0.058 | 2.180+/-0.123 | 12.3(1) | 0.107 | 2.18/2.65 |
| Sb1 | 6 | 0 | 1.989+/-0.004 | 10.4(1) | 0.02 | 5.17 |
| Sb2 | 6 | 0 | 2.033+/-0.005 | 11.0(1) | 0.003 | 5.21 |

Table 4. Crystallographic parameters of A$_3$BO$_6$ compounds with a corundum-related structure ($r_A$-ionic radius, lattice parameters $a$ and $c$, temperatures of magnetic (T$_N$ or Θ) from Curie-Weiss analysis and antiferroelectric (T$_C$) transition).

| Compound | $r_A$(Å) | $a$(Å) | $c$(Å) | T$_N$(K) | Θ(K) | T$_C$(K) |
|---|---|---|---|---|---|---|
| Mn$_3$TeO$_6$ | 0.83 | 8.8679(1) | 10.6727(2) | 24 | -120 | 510 |
| Mn$_2$InSbO$_6$ | 0.80 | 8.9313(1) | 10.7071(2) | - | -64 | 485 |
| Mn$_2$ScSbO$_6$ | 0.745 | 8.8836(1) | 10.6168(2) | - | -76 | 470 |

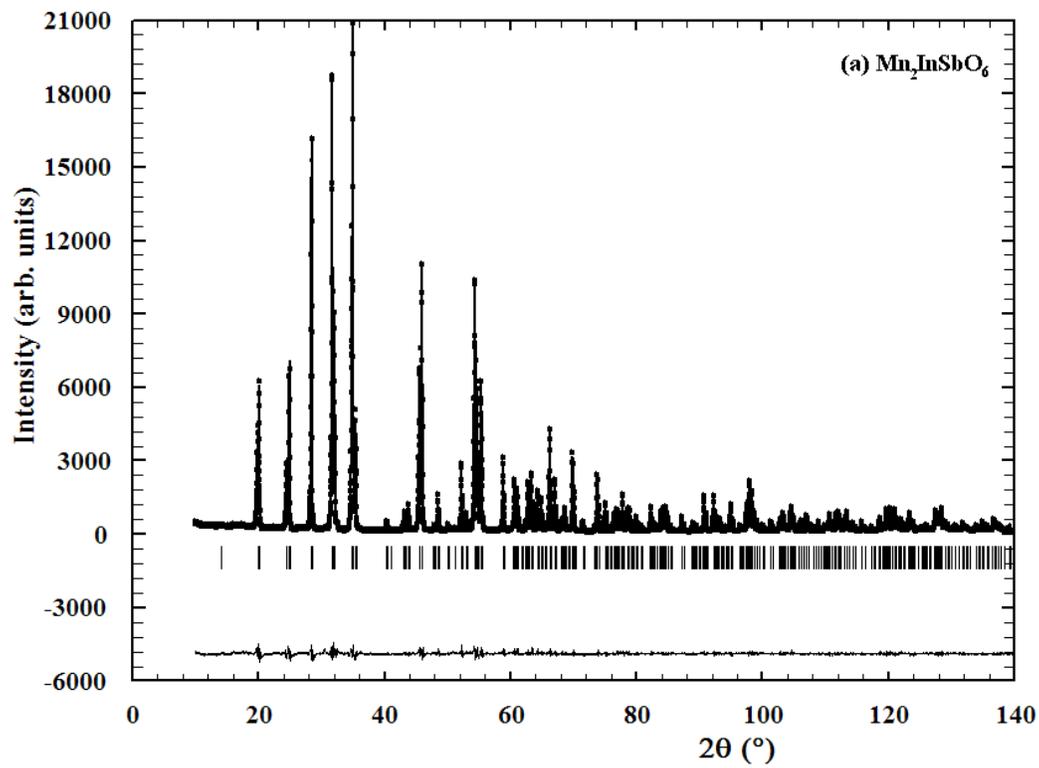

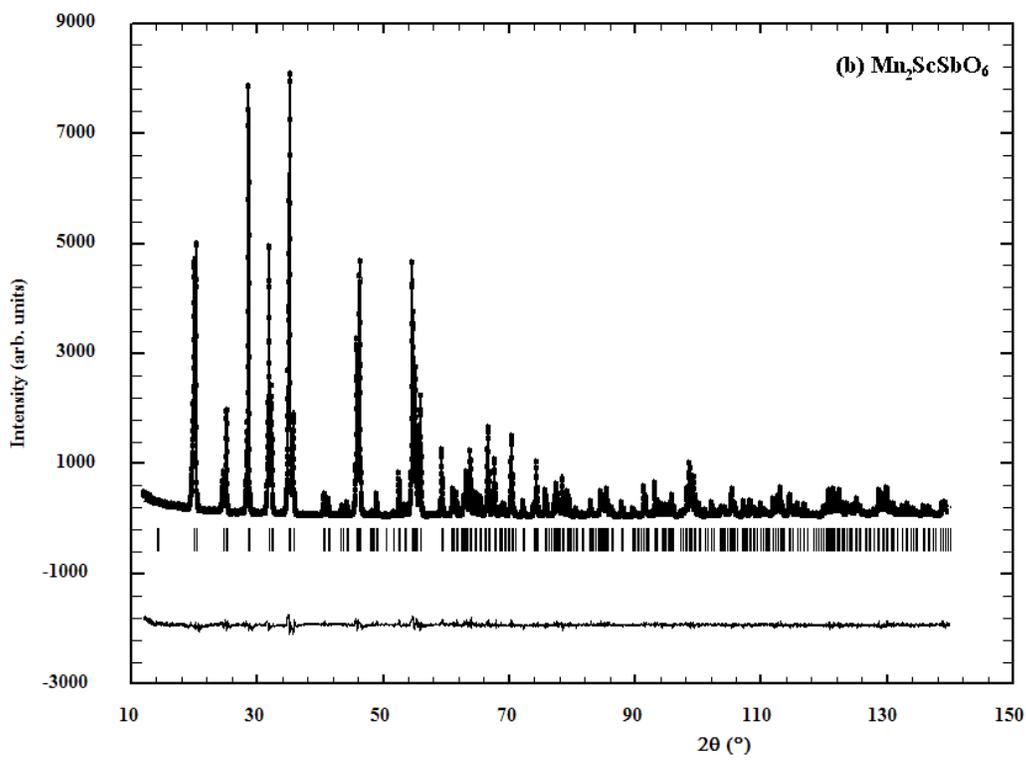

**Figure 1**

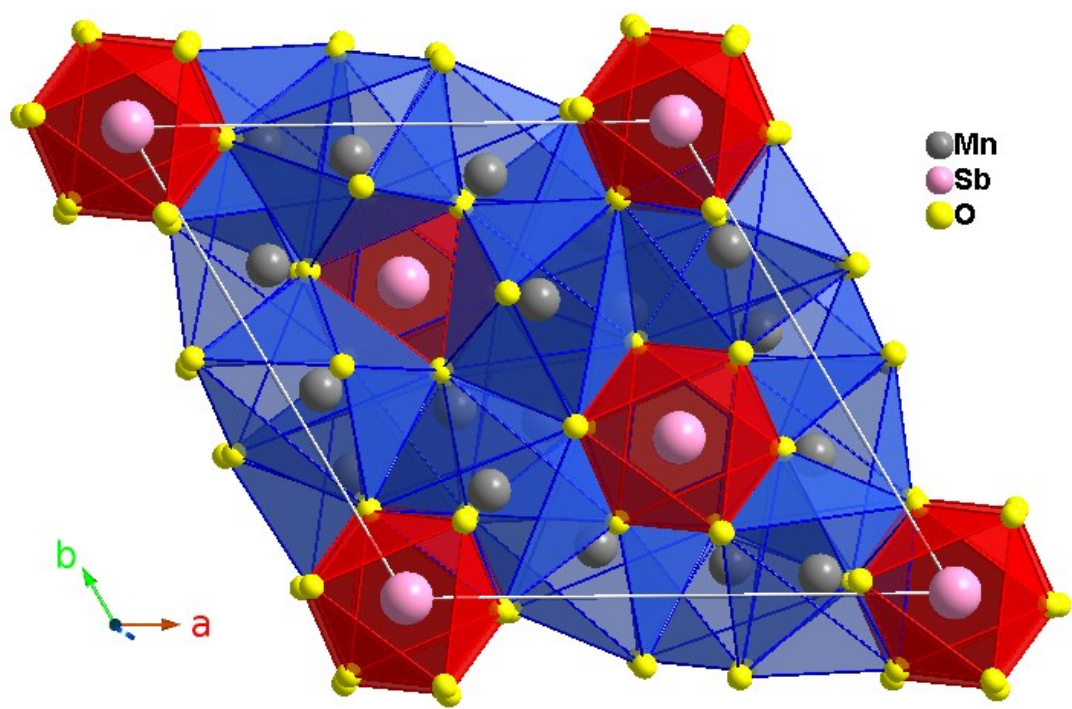

**Figure 2**

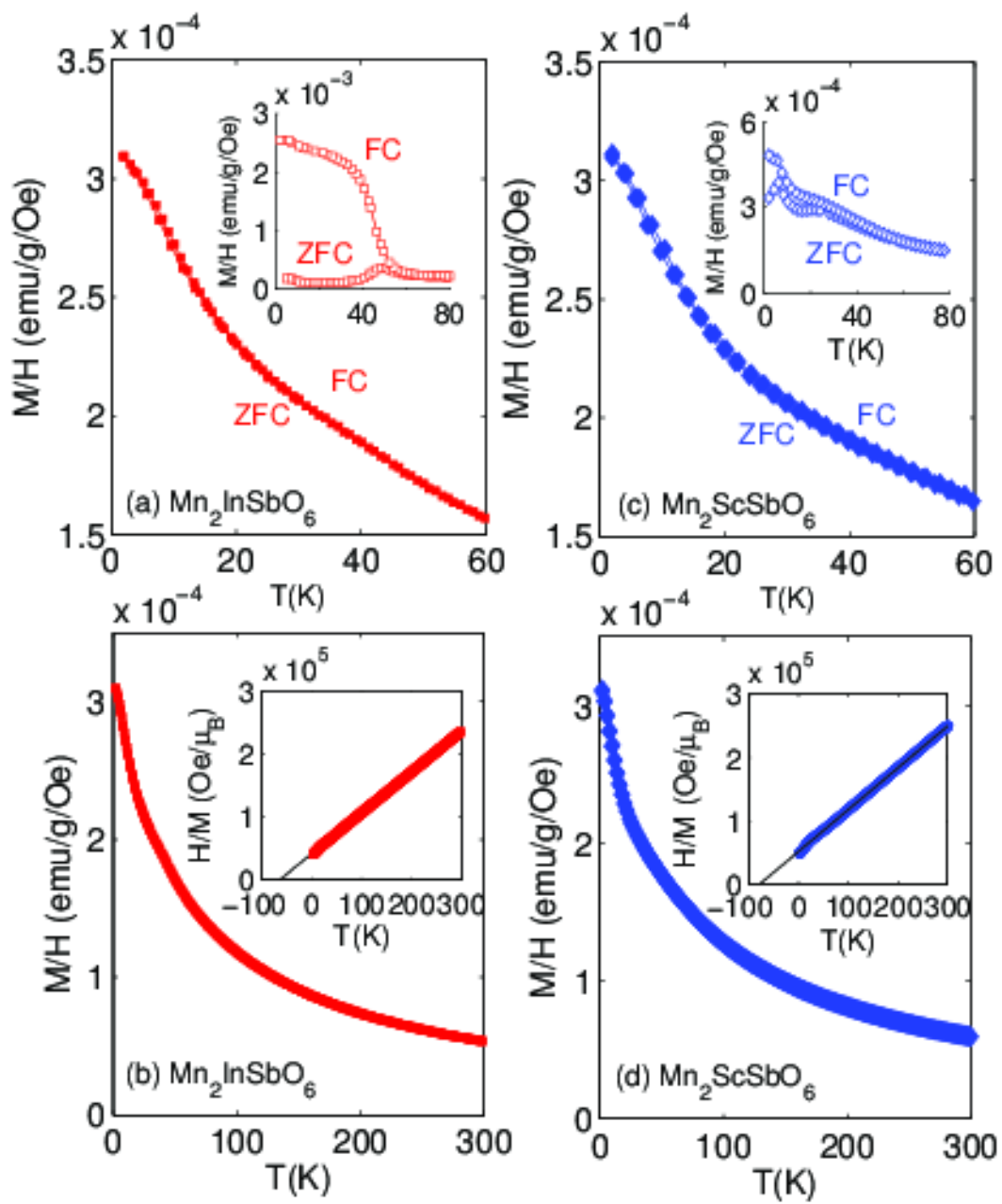

**Figure 3**

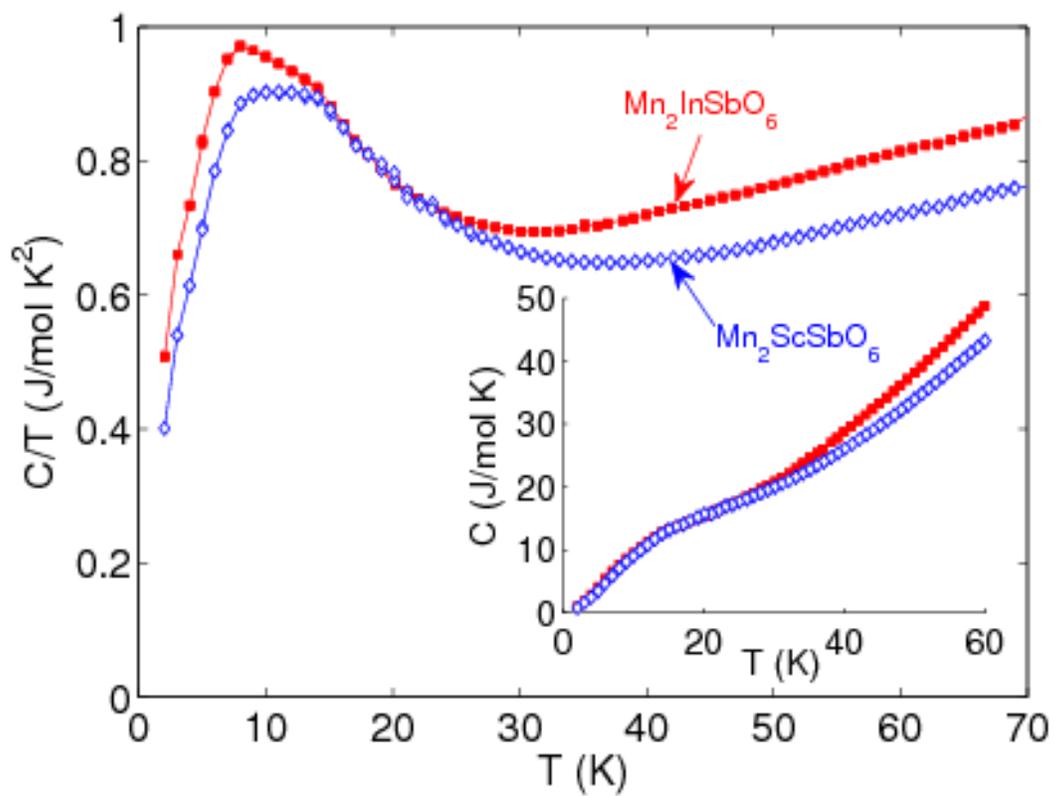

**Figure 4**

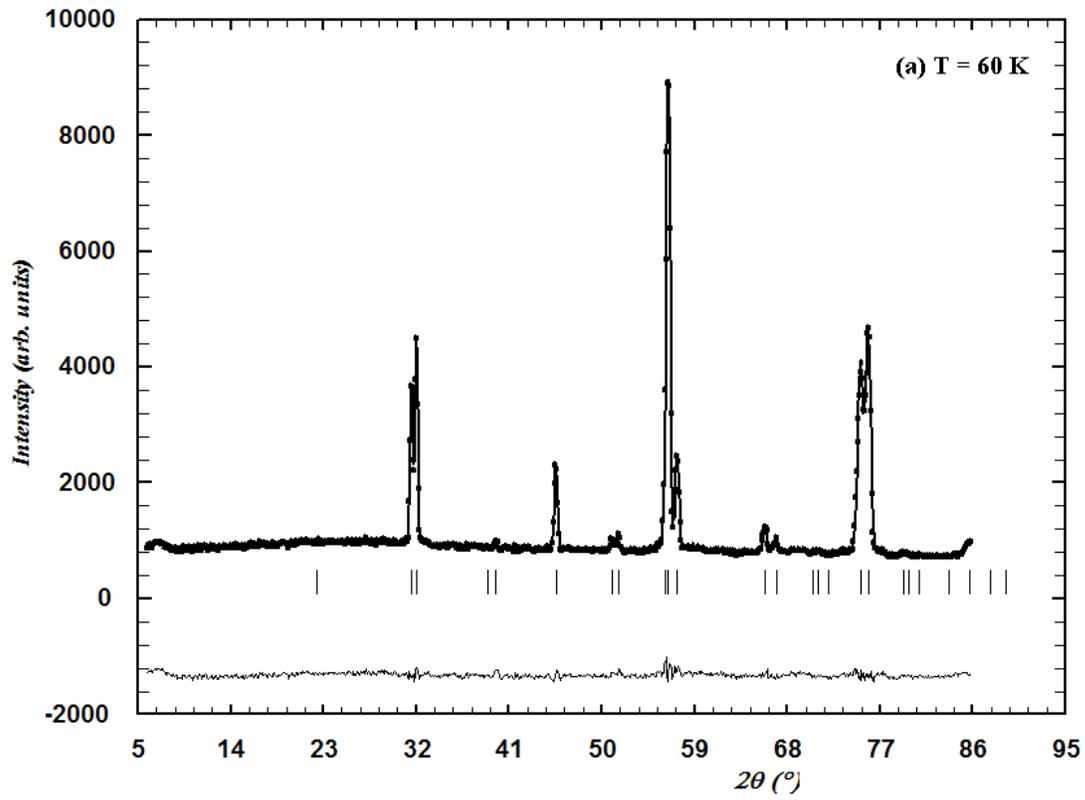

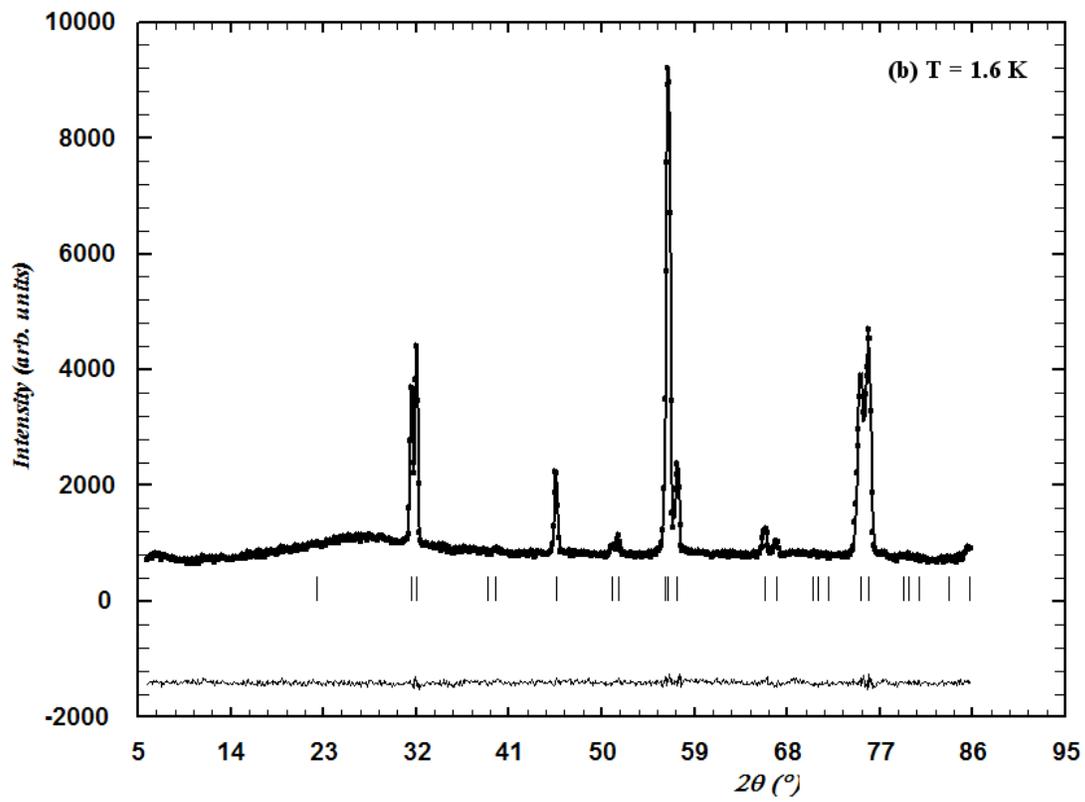

**Figure 5**

**List of captions for illustrations**

Figure 1. The observed, calculated, and difference plots for the fit to XRPD patterns of $Mn_2InSbO_6$ (a) and $Mn_2ScSbO_6$ (b) after the Rietveld refinement at 295 K.

Figure 2. A polyhedral view of the rhombohedral structure of $Mn_2InSbO_6$ at room temperature.

Figure 3. Upper panels: Temperature dependence of the ZFC/FC susceptibility M/H of (a) $Mn_2InSbO_6$ and (c) $Mn_2ScSbO_6$ in 3T (main frames) and 5 mT (insets). Lower panels: Temperature dependence of the ZFC/FC susceptibility M/H of (b) $Mn_2InSbO_6$ and (d) $Mn_2ScSbO_6$ in 3 Tesla up to higher temperatures. Associated Curie-Weiss fits are shown in the insets.

Figure 4. Temperature dependence of the heat capacity C (plotted as C/T in main frame and C in inset) of $Mn_2InSbO_6$ and $Mn_2ScSbO_6$

Figure 5. The observed, calculated, and difference plots for the fit to NPD patterns of $Mn_2InSbO_6$ after the Rietveld refinement at 60 K (a) and 1.6 K (b).

# Author's addresses

### Sergey A. Ivanov
Department of Inorganic Materials, Karpov' Institute of Physical Chemistry, Vorontsovo pole, 10 105064 Moscow K-64, Russia
**E-mail**: ivan@cc.nifhi.ac.ru, Sergey.Ivanov@mkem.uu.se

### Per Nordblad, Roland Mathieu
Department of Engineering Sciences, The Ångstrom Laboratory, Box 534, University of Uppsala, SE-751 21 Uppsala, Sweden
**E-mail**: Per.Nordblad@angstrom.uu.se

### Roland Tellgren
Department of Materials Chemistry, The Ångstrom Laboratory, Box 538, University of Uppsala, SE-751 21 Uppsala, Sweden
**E-mail**: Roland.Tellgren@mkem.uu.se